\documentstyle[11pt]{article}
\begin{document}
\title{Wormholes and Time-Machines}
\author{Frank Antonsen\\ with Karsten Bormann \\ Niels Bohr Institute \\ 
Blegdamsvej 17, DK-2100 Copenhagen \O, Denmark}
\date{}
\maketitle 
\begin{abstract}
It has been proposed that wormholes can be made to function
as time-machines. This opens up the question of whether this can be
accomodated within a self-consistent physics or not. In this contribution
we present some quantum mechanical considerations in this respect. 
\end{abstract}

\section{Introduction}
Consider a flat space-time in which two regions (mouth A and mouth B) 
are connected by a throught (a wormhole) and assume that 
the intrinsic length of this wormhole is small compared to the distance 
between mouth A and mouth B in the flat space-time.
It can be shown that this geometry can lead to the existence of            
closed time-like curves either by, at some period of time, accelerating       
one mouth of the wormhole relative to the other [1,3--5] or by placing the    
mouths in regions of differing gravitational potentials. An observer       
on a closed time-like curve can influence not only his own future but also    
his past; the wormhole thus can be made to function as a time machine.\\
The problem with which we are especially concerned is the ``quantum billiard
problem,'' the quantum analogue of the classical problem proposed by
Novikov et al. [2,4,13] The idea, in the classical case, is the following, consider
a wormhole acting like a time machine, denote the two mouths by A and B respectively.
We can assume that an object by traveling from A to B moves back in time by a
certain (fixed) timestep $T$ (as seen from the external essentially flat space-time). 
Now suppose we let a billiard ball roll towards
mouth A, it will then go through the wormhole and ``reappear'' in the past at
mouth B. Now, this gives the possibility that the ball exiting from mouth B
hits the original ball. Unless this scattering is sufficiently weak, the original
ball will be scattered off its original course and may not enter mouth A, in which case
it obviously cannot go back in time, and exit from B just in time to hit itself off course,
and thus it {\em will} reach mouth A... And we have a paradox. Novikov's idea was to
allow only self-consistent solutions, i.e. the ball exiting from mouth B is allowed
to hit the original ball on its way to mouth A, provided that it does not lead
to a radically new trajectory for the original ball. Whatever happens, the original ball
{\em must} reach mouth A. It has been proven that such selfconsistent solutions do exist
classically. The question is, do they also exist quantum mechanically? In the quantum
billiard problem, in an attempt to answer this, we consider wave packets instead of
balls, and these interact through some potential. Pertubation theory then gives us
a selfconsistency requirement, which we can actually solve in a few cases.\\
A related problem is the question of unitarity. In order to study this we must write
down a model Hamiltonian and calculate the evolution operator. For reasons of space, 
this line of attack will not be followed here, instead I just refer to $[16]$.\\
This seems to us to be the obvious way of attempting to answer these questions, but
unfortunately it is not without problems. A space-time
possesing closed time-like curves is not foliable, and hence we cannot, {\em 
in the vicinity of the wormhole}, make the 3+1 splitting of space-time 
essential to Schr\"{o}dinger mechanics.  This problem can be overcome when, 
instead of including the wormhole directly,
we include it only in an effective theory, in such a way that we do not have to use
Schr\"{o}dinger theory near the wormhole but only sufficiently far away from it
where we empirically know ordinary quantum theory to be correct. 
In this effective theory, the
possibility of going through the wormhole and backwards in time gives rise to an
interaction which looks much like an ordinary self-energy diagram. 
Away from the wormhole, Schr\"{o}dinger mechanics must be valid but
we have to take the self-interactions introduced by the presence of a time
machine into account.\\
To keep things so simple that an
analytical solution to the problem is possible we will assume
that the only effect of the wormhole on the wave packet, besides
moving it to a different place in space-time, is the possibility
of a shift in momentum. The diverging lense effect of the
wormhole [4] will be ignored, as will the scattering of the wave upon the
wormhole mouths. As we are thus only interested in the possibility or impossibility of
selfconsistent motion we think that this crude
model of the wormhole's interaction with its surroundings should suffice. 
Any quantum mechanical model taking the wormhole (and the
resulting absence of a foliation) into considerations
has to be equivalent to the Schr\"{o}dinger theory
sufficiently far away, where space-time is supposed to be flat. Hence
sufficiently far away, any model has to be equivalent to ours, although perhaps
with a different scattering kernel. If not, the mere applicability of
Schr\"{o}dinger quantum theory today would exclude the existence, anywhere in the
universe, of regions with closed time-like curves.

\section{The Selfconsistency Requirement for a Non-Relativistic Wave-Packet
with Coulomb Interactions in 2+1 Dimensions}
The simplest nom-trivial spatial dimension is two, and for simplicty we will
restrict ourselves to that; the results, however, would not be radically different
in higher dimensions.\\
From non-relativistic quantum mechanics, we know that a wave-packet $\psi_i$ can
undergo a transition $\psi_i\rightarrow \psi_f$ in the presence of a
pertubation. To first order, pertubation theory gives
\begin{equation}
    \psi_f({\bf k'}) = \int \frac{V_{\bf kk'}}{\epsilon_{\bf k'}-\epsilon_
    {\bf k}}\psi_i({\bf k}) d^2k
\end{equation}
where  $V_{\bf kk'} = \langle {\bf k'}|V|{\bf k}\rangle$ is the 
matrix element of the pertubation and where $\epsilon_{\bf k}$ is the energy, $\epsilon_{\bf k}=
k^2/2m$. We will choose units in which $m=1$.\\
The case of the quantum billiard problem is quite special: 
The potential in which $\psi_i$ scatters is derived
from its future ``self'', $\psi_f$, which gives rise to a ``charge distribution''
$\rho({\bf x},t) = |\psi_f({\bf x},t)|^2$, i.e. $V=V[\psi_f]$. We will choose a
potential of the form\footnote{One must either choose this prescription for
an almost Coulomb potential or screen the Coulomb potential in order
to obtain a finite theory. Some comments on the case of the screened 
potential (the Yukawa potential) will be made at the end of this section. } 
\begin{equation}
    V(r) = \alpha' \rho({\bf x},t) r^{\epsilon-1} =
    \alpha' |\psi_f({\bf x},t))|^2 r^{\epsilon-1} 
    \equiv v(r)|\psi_f({\bf x},t)|^2
\end{equation} 
where $\epsilon$ is taken to be small.
Denoting the Fourier coefficients of the original state by $a_{\bf k}$, 
and those of
the scattered state (that which goes through the wormhole and scatters its
former ``self'') by $c_{\bf k}$
\begin{eqnarray*}
      \psi_i({\bf x}) &=& \frac{1}{(2\pi)^{n/2}}\int a_{\bf k}e^{-i{\bf k\cdot 
      x}} d^2x\\
      \psi_f({\bf x}) &=& \frac{1}{(2\pi)^{n/2}}\int c_{\bf k}e^{-i{\bf k\cdot
      x}} d^2x
\end{eqnarray*}
we obtain upon insertion in eq(1), by using the Fourier convolution 
theorem twice and changing variables a couple of times
\begin{eqnarray}
    c_{\bf k'} &=& \alpha'\int \frac{a_{\bf k}}{\epsilon_{\bf k'}-\epsilon_{\bf
    k}}c_{\bf l}^*c_{\bf q-l}\|{\bf k-k'-q}\|^{\epsilon-2} d^2q d^2l d^2k\\
    &\equiv& \int c_{\bf p}^*\hat{X}^{\bf p,q}_{\bf k'}c_{\bf q} d^2p d^2q
\end{eqnarray}
where we have introduced a scattering kernel
\begin{eqnarray}
    \hat{X}^{\bf p,q}_{\bf k'} &\equiv& \int\frac{a_{\bf k}\tilde{v}({\bf l})}
    {\epsilon_{\bf k'}- \epsilon_{\bf k+l+p+q}}d^2k d^2l\\
    &=& 2\int\frac{a_{\bf k}\tilde{v}({\bf l})}{({\bf k'})^2-({\bf k+l+p+q})^2}
    d^2kd^2l
\end{eqnarray}
where $\tilde{v}(k)=\alpha' \|k\|^{\epsilon-2}$ 
is the Fourier transform of $v(r)=\alpha'r^{\epsilon-1}$ and where we have
inserted $\epsilon_{\bf k} = \frac{1}{2}{\bf k}^2$. To find the
self-consistent solutions we must solve the (infinite dimensional)
quadratic equation (4). 
In a more general set-up the kernel would also contain information about the
structure and geometry of the wormhole. Thus it is essentially this quantity which
hides our ignorance of the detailed structure of the wormhole.

\subsection{On the Consistency and Limitations of this Formalism}
Now, as mentioned in the introduction, space-times with closed time-like curves
do not admit a foliation, and hence ordinary quantum mechanics is in principle
meaningless.
Thus, some comments on the consistency of the proposed formalism are in order.
First of all, we do not attempt to give a quantum description of the dynamics
close to the wormhole mouths. Sufficiently far outside 
that region, where space-time is almost flat, foliations are possible 
and ordinary
quantum mechanics is known to be valid. But we are interested in the interaction of
this ``forbidden region'' with its surroundings, as we want the wave-packet to traverse the wormhole.
We then try to set-up an {\em effective} theory, which can accomodate this. The
``forbidden region'' is considered as a kind of ``black box'', which interacts
with the environment: particles can enter it, and it emits particles too. This
would be completely analogous to the situation of a quantum mechanical system
interacting with a classical system, were it not for the added feature of
special (temporal) correlations. A particle entering the region at time 
$t$ is correlated
with a particle exiting at the earlier time $t-T$, where $T$ is the typical
time-step of the wormhole. If the original wave packet is to interact with it we
can desrcibe this in terms of an effective interaction $V=v(r)
|\psi_f({\bf x},t)|^2$, where $v(r)$ is the potential between two classical
point particles and $|\psi_f |^2$ is the charge-distribution.
We can reexpress the potential in terms of the initial
wave-packet by writing $|\psi_f({\bf x},t)|^2=w({\bf x},t)|\psi_i({\bf x},t)|^2$
whereby the effective potential becomes $V({\bf x},t) = w({\bf x},t)|\psi_i({\bf
x},t)|^2$. This holds provided we stay away from the at most a countable
number of zeroes of $\psi_i$, which thus form a set of measure zero.
Simply plugging this into a Schr\"{o}dinger equation leads to the following
effective equation of motion (dropping the subscript $i$ on $\psi$):
\begin{equation}
    -\frac{1}{2}\nabla^2\psi + w({\bf x},t)|\psi|^2\psi = 
    i\frac{\partial}{\partial t}\psi
\end{equation}
which is a slight generalization of the well-known non-linear Schr\"{o}dinger
equation [14], the only new feature being the non-constant coefficient $w({\bf
x},t)$.
\\
Giving up describing the dynamics inside the ``forbidden region'', we can
essentially use ordinary Schr\"{o}dinger mechanics outside, but with an
effective potential depending upon the wave function, thus leading to a
generalization of the non-linear Schr\"{o}dinger equation as the effective
equation of motion.
Hence, as an effective theory, the proposed formalism should suffice.
Thus also, in principle, one could calculate all sorts of transition amplitudes
using the scattering theory of this generalization of the non-linear 
Schr\"{o}dinger equation.\footnote{We are of course aware of the somewhat
hand-waving charachter of these arguments, but we do believe they can be made
more rigorous with very little effort. For simplicity, however, we have decided
to stick to this simplistic argument.}

\subsection{Gaussian Wave Packets in an Almost Coulomb Potential}
With a Gaussian wave-packet (the prototype of a localized wave-packet, i.e. the best
analogue of a classical object) as our initial wave function, parameterized as
\begin{equation}
    a_{\bf k} = e^{-ak^2+{\bf b\cdot k}+c}
\end{equation}
we obtain after a lenghty and tedious but standard calculation:
\begin{equation}
    \hat{X}^{\bf k'}_{\bf pq} = 2 \alpha'\pi^2 B(\epsilon,1-\epsilon)
    \int_0^\infty e^{-ax^2+c'}I_0(b'x)\frac{x}{(k^{'2}-x^2)^{1-\epsilon}} dx
\end{equation}
where $B(x,y)\equiv \frac{\Gamma(x)\Gamma(y)}{\Gamma(x+y)}$ is the Beta 
function, $I_0$ is a modified Bessel function, and the coefficients are
\begin{eqnarray}
    b' &=& \| {\bf b}-2a({\bf p-q})\|\\
    c' &=& c-a({\bf p+q})^2-{\bf b}\cdot({\bf p+q})
\end{eqnarray}
Proceed from eq(9) by noting that
the Beta function, $B$, is singular in the limit $\epsilon\rightarrow 0$.
This comes as no surprise as this is the place
where we put the infinities arising from the nature of the Coulomb field. 
Imagining the Beta function removed (or rather regularized) by renormalization
makes it plausible to put $\epsilon\equiv 0$
inside the integral and simply treat the Beta function taken at $\epsilon=0$
as a (finite) constant. Doing this we can evaluate the above integral obtaining
\begin{equation}
    \hat{X}^{\bf k'}_{\bf pq} = 2i\alpha'\pi^3B(\epsilon,1-\epsilon) 
    e^{-a(k')^2+c'}I_0(b'k')
\end{equation}
Note from eqs(4),(12) that $c_{\bf k'}$ will always go like a Gaussian times 
some function, i.e.
\begin{equation}
    c_{\bf p} = f({\bf p}) e^{-\alpha p^2}
\end{equation}
The selfconsistency condition then reads 
\begin{eqnarray}
    c_{\bf k'} = &&2\pi\xi e^{-ak^{'2}}\sum_{nm}b_nb_m\int_0^\infty 
    (p_+^2-p_-^2)^
    {\frac{n+m}{2}}P_{\frac{n+m}{2}}\left(\frac{p_+^2+p_-^2}{p_+^2-p_-^2}\right)
    \times\nonumber\\
    &&\qquad 2^{(n+m)/2}I_0(Ap_-)e^{-\alpha p_-^2-(a+\alpha)p_+^2}p_+p_-dp_+dp_-
\end{eqnarray}
where we have defined
\begin{equation}
    A = 2^{3/4}\sqrt{a}k'\mbox{     and    } \xi = 2i\alpha'\pi^3B(\epsilon,
    1-\epsilon)e^c
\end{equation}
with ${\bf p}_\pm = 2^{-1/2}({\bf p}\pm {\bf q})$ and where $b_n$ denotes the
Taylor coefficients of $f({\bf p})=\sum b_p{\bf p}^n$ (one can prove that only
even powers of $\bf p$ can possibly satisfy the selfonsistency requirement).
This integral can actually be carried out, and we arrive at
\begin{eqnarray}
    c_{\bf k'} &=& e^{-\alpha{\bf k'}^2}\sum_n b_{2n}{\bf k'}^{2n}\nonumber\\ 
    &=&2\pi^2\xi e^{-ak^{'2}}\sum_{mn}\left[b_{2n}b_{2m}\sum_{kl}
    \left(\begin{array}{c}n\\k\end{array}\right)\left(\begin{array}{c}m\\l
    \end{array}\right)2^{k+l+n+m}\frac{(k+l-1)!!}{(k+l)!!}\times\right.\nonumber\\
    &&\hspace{-1cm}\left.\sum_{l'} \left(
    \begin{array}{c} n+m-k-l\\l'\end{array}\right)\frac{\Gamma(\frac{n+m+2-l'}
    {2})}{\alpha^{(n+m+2-l')/2}}\Phi(\frac{n+m+2-l'}{2},1;\frac{A^2}{4\alpha})
    C_{k+l+l'+1}\right]\nonumber\\
\end{eqnarray}
where $\Phi(a,b;z)$ is a degenerate hypergeometric function [6]
\begin{displaymath}
    \Phi(a,b;z) \equiv \sum_{n=0}^\infty \frac{(a)_n z^n}{(b)_n n!}
\end{displaymath}
with $(a)_n \equiv a(a+1)(a+2)...(a+n-1)$, and where we have introduced
\begin{equation}
    C_\nu \equiv \left\{\begin{array}{ll}
    \frac{(2\lambda-1)!!}{2(2(a+\alpha))^\lambda} \sqrt{\frac{\pi}{a+\alpha}
    } & \nu=2\lambda\\
    \frac{\lambda!}{2(a+\alpha)^{\lambda+1}} & \nu=2\lambda+1
    \end{array}\right.
\end{equation}
Note that this is an exact result, no approximations have been made.
This equation constitutes the {\em final form of the selfconsistency 
requirement} in the case of an incomming Gaussian wave packet interacting through
a Coulomb potential, and so it is this equation we have to solve to find
self consistent solutions. This selfconsistency requirement can 
only be solved (in principle) in the two extreme
cases: $f={\rm constant}$ and $f$ not a polynomial (i.e. the Taylor series never
terminates, in which case $f$ would be some analytic function of $k^{'2}$). 
Due to the hypergeometric function on the right hand side, 
the selfconsistency equation has no solutions when $f$ is a polynomial. \\
If the wave packet is a pure Gaussian, we get a
requirement on the wave packet traversing the wormhole (remember that the 
$\alpha$ refers to $\psi_f$, whereas the $a$ refers to $\psi_i$). By 
putting $c_{\bf k'} =b_0 \exp(-\alpha k^{'2})$  (cf.eq(12)) on the left hand 
side and similarly on the right hand
side, where only one term in the sum would then appear, one immediatly sees that
\begin{equation}
    b_0 = b_0^2 2\pi^2\xi
\end{equation}
i.e. $b_0= (2\pi^2\xi)^{-1}$ (or $b_0=0$, but this would correspond to $\psi 
\equiv 0$ and is hence not interesting). Also, by differentiating twice with 
respect to $k'$ and putting $k'=0$ one gets
\begin{equation}
    \alpha = \frac{1}{2}a \pm \frac{1}{2}\sqrt{a^2+2\sqrt{2}a}
\end{equation}
i.e. there are exactly two solutions for the scattered wave packet. \\
This case, where the wave-packet is Gaussian at all times (i.e. both before 
and after scattering) is the quantum 
analogue of the classical billiard problem (the Gaussian wave-packets
which are substituted for the billiard balls are as localized as the
Heisenberg uncertainty principle allows).\\
The normalization of the wave-packet introduces a
requirement on the {\em original} wave packet (i.e. on $a$). 
Normalization of 
the Gaussian wave-packet would require $b_0=\sqrt{\frac{2\alpha}{\pi}}$, 
i.e. in order to have normalized wave packets we would have to impose 
the requirement
\begin{equation}
    \alpha = (8\pi^3\xi^2)^{-1} = -(64ia\pi^7\alpha^{'2}B^2(\epsilon,1-
    \epsilon))^{-1}
\end{equation}
where we have inserted the definition of $\xi$ (see eq(14), also, 
remember that $\alpha'$ is the coupling constant) and demanded that the original
wave-packet is normalized (i.e. $e^c=\sqrt{2a/\pi}$). This then allows only 
one {\em original Gaussian}, namely that with $a$ satisfying
\begin{equation}
    a^2\pm a\sqrt{a^2+2\sqrt{2}a} = -(32i\pi^7\alpha^{'2}B^2
    (\epsilon,1-\epsilon))^{-1}
\end{equation}
Thus there are only two solutions; in units where the right hand side is equal to one,
we find $a=0.5337543$ when we use the plus sign and $a=1.4799995$ when we use 
the minus sign.\\
Had we chosen a Yukawa potential, $v(r)=\alpha'r^{-1}e^{-r/m}$, instead
we would have had to make the substitution $k^{'2}\rightarrow k^{'2}+m^2$
in all the expressions, and the coefficients would become non-singular
in the limit $\epsilon\rightarrow 0$. Hence eq(12) would become
\begin{displaymath}
    \hat{X}^{\bf k'}_{\bf pq} \propto  e^{-a(k^{'^2}+m^2)+c'}I_0(b'\sqrt{k^{'2}
    +m^2})
\end{displaymath}
and the quantity $A$ defined in eq(15) would become $A=2^{3/4}\sqrt{a} 
\sqrt{k^{'2}+m^2}$. This would make finding solutions in the general case even
more difficult, as the right hand side of the selfconsistency requirement,
eq(16), now would contain terms of the form $(\sqrt{k^{'2}+m^2})^n$ where
$n$ is just some integer. This will make analytical solutions very difficult to find.
The Gaussian solution would still exist, though, but
with $b_0$ multiplied by $\exp(am^2)$. Similarly, the right hand side of eq(20)
would also change, but can still be taken to unity by an appropriate choice
of $m$ and $\alpha'$. One can furthermore also consider plane waves, these do not,
however, correspond to classical billiard balls, and moreover, analytical solutions
are much more difficult to find.

\subsection{Stability of Solutions}
In the preceding section we found solutions to the selfconsistency requirement
for pure Gaussians. By analogy with the classical billiard problem, where one
attempts to avoid the situation in which the scattering of the incoming ball 
on its
future self makes the incoming ball fail to traverse the wormhole, we now 
want to investigate the stability of these solutions under small perturbations. 
Write the selfconsistency condition in a symbolic form as
\begin{equation}
    c_{\bf k'} = \hat{X}^{\bf p,q}_{\bf k'}c_{\bf p}c_{\bf q}
\end{equation}
invoking a generalized summation convention consisting in integrating over
repeated indices. We then consider a slight pertubation $\delta c_{\bf k}$ of a
fixed solution $\zeta_{\bf k}$, to first order in the pertubation we then get
\begin{equation}
    \delta c_{\bf k'} = \hat{X}^{\bf p,q}_{\bf k'}\zeta_{\bf p}\delta c_{\bf q}
    +\hat{X}^{\bf p,q}_{\bf k'}\zeta_{\bf q}\delta c_{\bf p} \equiv 
      \hat{M}_{\bf k'}^{~\bf l}\delta c_{\bf l}
\end{equation}
where
\begin{equation}
    \hat{M}_{\bf k'}^{~\bf l} \equiv \hat{X}^{\bf p,q}_{\bf k'}\left(
      \zeta_{\bf p}
    \delta_{\bf q}^{~\bf l}+\zeta_{\bf q}\delta_{\bf p}^{~\bf l}\right)
\end{equation}
with $\delta_{\bf k}^{~\bf l}\equiv\delta^{(n)}({\bf k-l})$.\\
To study the stability of the solution $\zeta_{\bf p}$ we must consider the 
infinite dimensional map 
\begin{equation}
    \delta c_{\bf k'}^{(n)} \mapsto
    \delta c_{\bf k'}^{(n+1)} = \hat{M}_{\bf k'}^{~\bf l}\delta c_{\bf l}^{(n)}
\end{equation}
The difference, $\Delta_{\bf k}^{(n)}$, between two such iterates is then simply
\begin{equation}
    \Delta_{\bf k'}^{(n)} = |(\hat{M}_{\bf k'}^{~\bf l}-\delta_{\bf k'}^{~\bf 
      l}) 
    \delta c_{\bf l}^{(n)}| = |(\hat{M}_{\bf k'}^{~\bf l}-\delta_{\bf k'}
      ^{~\bf l})^n\delta c_{\bf l}^{(0)}| 
\end{equation}
This difference then goes like the size of the eigenvalue of $\hat{M}_{\bf k'}
^{~\bf l}$ corresponding to $\bf k'$. Denoting this eigenvalue by 
$\lambda({\bf k'})$ we get
\begin{equation}
    \Delta_{\bf k'}^{(n)} \sim |(\lambda({\bf k'})-1)^n||\delta 
      c_{\bf k'}^{(0)}|
\end{equation}
i.e. it diverges when $\lambda({\bf k'})> 2$,in which case then, the solution is
unstable against slight pertubations. If, on the other hand, $\lambda({\bf k'}) 
< 2$ then the corresponding solution $\zeta_{\bf k}$ is stable. In accordance 
with the language of chaos theory we call $\lambda({\bf k'})$ the {\em 
generalized Lyapunov exponent}.\footnote{In the usual finite dimensional case 
treated in chaos theory, this exponential is a function of the solution 
$\vec{\zeta}$, which in our infinite dimensional analogue means that 
$\lambda$ is a functional of $\zeta_{\bf k}$.}\\
We have reduced the problem of stability to that of finding the eigenvalues of
the integral operator $\hat{M}_{\bf k}^{~\bf l}$.\\
Now proceed to examine the pure Gaussian solution found in the 
previous section;
\begin{equation}
    \zeta_{\bf k} = b_0 e^{-\alpha k^2} \qquad (=c_{\bf k})
\end{equation}
Inserting this into the definition of the operator $\hat{M}_{\bf k}^{~\bf l}$ we
get, by construction, essentialy the same integrals as those we performed 
in order to solve the selfconsistency requirement. Explicitly
\begin{equation}
    \lambda({\bf k}) = \left(b_0e^{-a k^2}\left(\int I_0(k\|{\bf p-l}\|) 
      e^{-a({\bf p+l})^2-\alpha p^2}d^2p + ({\bf p}\rightarrow{\bf q})
      \right)\right)\delta_{\bf k}^{~\bf l}
\end{equation}
We can get, by ignoring the terms linear in ${\bf p,q}$ in the exponent, an 
approximate expression for these integrals
\begin{equation}
    \lambda({\bf k}) \approx 2be^{-\left(2a+\frac{1}{8(a+\alpha)^2}\right)k^2}
    \sqrt{\frac{\pi}{a+\alpha}}I_0\left(\frac{k^2}{8(a+\alpha)^2}\right)
\end{equation}
Into this we can then insert $b_0=\sqrt{\pi/\alpha}$ and $\alpha=(a+\sqrt{a^2+2a\sqrt{2}})/2$.
Doing this we find a stable ($\lambda < 2$) as well as 
an unstable ($\lambda > 2$) region. It turns out that the solution is unstable 
for small $a$, but gets more and more stable as $a$ grows, i.e. as the {\em 
original} wave packet becomes more and more localized in momentum space. But 
this implies that the wave packets are very diffuse in position space, i.e. 
does not at all look like a classical point particle. In fact, the more the 
quantum nature is apparent i.e. the larger the uncertainty in position, the 
better the stability of the Gaussian solution. We conclude the selfconsistent
solutions of the classical billiard problem of e.g. Novikov [2] in this quantum mechanical
framework (localized wave-packets) are unstable.

\section{Discussion \& Conclusion}
One should first of all notice that we did not use any knowledge of the
wormhole, nor did we specify where the interactions take place; this just has to
be sufficiently (depending on the size of the wormhole) far away from it, where
space-time is flat. Only the momentum of the packets were specified, and hence
their location is indeterminate -- the wormhole simply effectively introduces a new
(self) interaction.\\
We ignored any effect the traversal of the wormhole
might have on the wave packet except for a possible shift
in momentum. In particular we let the diverging lense 
effect [4] out of consideration and the scattering of the wave upon the mouths.
This does not seem to us 
to be able to alter the conclusions of this paper, because the
resulting smaller amplitude, and consequently smaller scattering,
could be compensated by changing the geometry of the problem, i.e.
by changing the distance between the `out-mouth' and the region 
where scattering occurs. Also of course, the assumption that 
the wave packets be small as compared to the wormhole is
essential to the calculations made.\\
We have also ignored
the possibility of the wave packet going through the wormhole
more than once and therefore getting a larger shift in time (but with smaller
amplitude of the shifted wave, due to the diverging lense effect),
because this would just alter the region in which the scattering occurs; 
an effect which could be compensated again by changing the geometry 
appropriately. By the same token, we let out of consideration the possibility 
of writing the wave after scattering as a superposition of waves that have 
traversed the wormhole a different number of times. This problem probably could
be treated in a second quantized version of the above model, but could 
also be seen as just going to higher orders in the perturbation 
expansion and probably would not change much -- it should be equivalent to a
proper path-integration approach with a suitable highly non-trivial measure,
taking only self-consistent solutions into account.\\
Thus we almost completely ignored the wormhole, which was the reason why we
could use a Hamiltonian formulation. We have only included the time-machine
effect of the wormhole, and this in a rather indirect manner, through an
effective potential and hence through an effective equation of motion. This
equation of motion turned out to be essentially the non-linear Schr\"{o}dinger
equation.
\\
We derived a general, closed equation expressing the requirement of
selfconsistency. This equation could be solved exactly only in the
case where the Fourier coefficents of the wave packet after scattering, i.e. 
the part of the wave packet traveling on the closed timelike curve, has the form
$c_{\bf k'} =b_0 \exp(-\alpha k^{'2})$. If the solution
was normalized we found only one possible value of the width of the
incoming wavepacket and that the corresponding solution was unstable in 
large parts of parameter space, so only
fine-tuning of initial conditions could render physics selfconsistent 
in these parts of parameter space. This need for fine-tuning springs from the
restrictions on the form of the wave-packet. If one threw away this requirement
the need for fine-tuning, i.e. the restrictions imposed upon the initial
conditions, would in all likelihood become very much less severe. On the other
hand, this form-requirement was essential for a semi-classical picture of
``billiard balls'' self-interacting due to the presence of a time-machine.\\
A fuller discussion of these problems will be given in $[15-16]$.
                                                                    
\section*{References}
$[1]$ F.Echeverria, G.Klinkhammer \& K.Thorne: Phys. Rev. {\bf D44}(1991)1077\\
$[2]$ I.D.Novikov: Phys. Rev. {\bf D45}(1992)1989\\
$[3]$ M.S.Morris, K.S.Thorne, U.Yurtsever: Phys.Rev.Lett.{\bf D61}(1988)1446\\
$[4]$ S.-W.Kim, K.S.Thorne: Phys.Rev.{\bf D43}(1991)3929\\
$[5]$ J.Friedman {\em et al.}: Phys.Rev.{\bf D42}(1990)1915\\
$[6]$ I.S.Gradshteyn \& I.M.Ryzhik: Table of Integrals, Series and Products 
(Academic Press, New York 1980)\\
$[7]$ G.Klinkhammer \& K.Thorne: unpublished preprint alluded to in $[1]$\\
$[8]$ D.Deutsch: Phys.Rev.{\bf D44}(1991)3197.\\
$[9]$ J.Friedman {\em et al.}: Phys.Rev.{\bf D46}(1992)4456.\\
$[10]$ D.G.Boulware: Phys.Rev.{\bf D46}(1992)4421.\\
$[11]$ H.D.Politzer: Phys.Rev.{\bf D46}(1992)4470.\\
$[12]$ J.Hartle: LANL bulletin board gr-qc9309012.\\
$[13]$ A.Lossev \& I.D.Novikov: (preprint) Nordita-91/41 A.\\
$[14]$ T.Taniuti \& N. Yajima: J.Math.Phys.{\bf 10}(1969)1369. \\
$[15]$ F.Antonsen, K.Bormann: ``The Self-Consistency of the Quantum Billiard Problem in Wormhole
Space-Times'' (to be submitted to Int.J.Theor.Phys.)\\
$[16]$ F.Antonsen, K.Bormann: ``Time-Machines and the Breakdown of Unitarity'' (to be submitted to Int.J.Theor.Phys.)
\end{document}